\documentstyle[12pt]{article}

\newcommand{\nc}{\newcommand}
\def\frac#1#2{{\textstyle {#1 \over #2}}}

\nc{\beq}{\begin{equation}}
\nc{\eeq}{\end{equation}}
\nc{\beqa}{\begin{eqnarray}}
\nc{\eeqa}{\end{eqnarray}}
\nc{\lsim}{\begin{array}{c}\,\sim\vspace{-21pt}\\< \end{array}}
\nc{\gsim}{\begin{array}{c}\sim\vspace{-21pt}\\> \end{array}}

\def\mev{\rm\,MeV}

\def\&{and}

\def\nc#1#2#3{           {\it Nuovo Cim.  }{\bf #1}, #2 (19#3)}
\def\np#1#2#3{           {\it Nucl. Phys. }{\bf #1}, #2 (19#3)}
\def\pl#1#2#3{           {\it Phys. Lett. }{\bf #1}, #2 (19#3)}
\def\pr#1#2#3{           {\it Phys. Rev. }{\bf #1}, #2 (19#3)}
\def\prep#1#2#3{         {\it Phys. Rep. }{\bf #1}, #2 (19#3)}
\def\prl#1#2#3{          {\it Phys. Rev. Lett. }{\bf #1}, #2 (19#3)}

\begin{document}

\begin{titlepage}

\begin{center}
     \hfill       YCTP-P14-96\\
\vskip .5 in
{\Large \bf  QCD at Large $\theta$ Angle and
Axion Cosmology
}
\vskip .3 in


{{ Nick Evans\footnote{nick@zen.physics.yale.edu},
Stephen D.H. Hsu\footnote{hsu@hsunext.physics.yale.edu},
Andreas Nyffeler \footnote{nyffeler@genesis2.physics.yale.edu}   and Myckola
Schwetz\footnote{ms@genesis2.physics.yale.edu}
}}

   \vskip 0.3 cm
   {\it Sloane Physics Laboratory,
        Yale University,
        New Haven, CT 06520}\\
\end{center}

\vskip .5 in
\begin{abstract}
\noindent
We use the chiral Lagrangian to investigate the global
properties of the $N$-flavor QCD vacuum as a function of the
$\theta$ parameter. In the case of exact quark degeneracy
we find evidence for first order
phase transitions at $\theta = \pi \cdot ({\rm odd~ integer} ) $.
The first order transitions are smoothed by quark mass
splittings, although interesting effects remain at
realistic quark masses. We emphasize the role of the
$\eta'$ condensate in our analysis.
Finally, we discuss the implications of
our results on the internal hadronic structure of axion
domain walls and axion cosmology.

\end{abstract}
\end{titlepage}

\input epsf
\newwrite\ffile\global\newcount\figno \global\figno=1
\def\writedefs{\immediate\openout\lfile=labeldefs.tmp \def\writedef##1{%
\immediate\write\lfile{\string\def\string##1\rightbracket}}}
\def\writestoppt{}\def\writedef#1{}

\def\figin{\epsfcheck\figin}\def\figins{\epsfcheck\figins}
\def\epsfcheck{\ifx\epsfbox\UnDeFiNeD
\message{(NO epsf.tex, FIGURES WILL BE IGNORED)}
\gdef\figin##1{\vskip2in}\gdef\figins##1{\hskip.5in}
\else\message{(FIGURES WILL BE INCLUDED)}%
\gdef\figin##1{##1}\gdef\figins##1{##1}\fi}

\def\figinsert{}
\def\ifig#1#2#3{\xdef#1{fig.~\the\figno}
\writedef{#1\leftbracket fig.\noexpand~\the\figno}%
\figinsert\figin{\centerline{#3}}\medskip\centerline{\vbox{\baselineskip12pt
\advance\hsize by -1truein\center\footnotesize{  Fig.~\the\figno.} #2}}
\bigskip\endinsert\global\advance\figno by1}
\def\footnotefont{}\def\endinsert{}

\renewcommand{\thepage}{\arabic{page}}
\setcounter{page}{1}

\section{Introduction}

The QCD Lagrangian may be augmented with the gauge invariant term
$\theta~ G\tilde{G}/32 \pi^2$ which,
despite being a total divergence, can have physical
effects due to gauge field configurations with non-trivial topology.
In nature
this term appears to be absent (or at least small -- $\theta \leq 10^{-10}$
\cite{CP}),
 but it is of theoretical
interest to determine the effect such
a term has on the theory.  It is possible that understanding the effects of a
non-zero $\theta$ angle
may help determine why the term is absent in the strong interactions. In this
paper we shall
study the effects of adding a non-zero $\theta$ angle on the QCD vacuum
assuming it
does not change the pattern of chiral symmetry
breaking ($SU(N)_L \otimes SU(N)_R \rightarrow SU(N)_V$)
in the quark sector.
While it is possible that adding a non-zero theta term to QCD may have more
dramatic effects, such as altering the nature of confinement \cite{TH,SCHZ},
we consider ours to be the more conservative assumption.
In fact,  it can be shown \cite{us} that some exactly
solved softly broken supersymmetric gauge theories
which confine by dyon condensation
continue to do so for all $\theta$ \footnote{However,
in these models the low energy effective $\theta$ angle
depends in a complicated fashion on the bare $\theta$ angle.}.
If the behaviour of QCD is smooth with changing $\theta$
then the low energy theory may be described
in terms of the standard chiral Lagrangian formalism. Witten \cite{Witten}
has previously studied this
possibility and shown that there may be first order phase transitions at
$\theta = \pi$. He
was particularly interested in the possibility of spontaneous CP violation at
$\theta = \pi$ (note that although $\theta \neq 0$ in general violates CP,
$\theta = \pi$ actually
conserves CP). Creutz \cite{Creutz}
further explored the phase structure of the mass degenerate
chiral Lagrangian, demonstrating the existence of  first order phase
transitions at
$\theta = n \pi$ ($n \in$ odd integers).
Here we
develop these observations with further study of these phase
transitions, concentrating in particular on the role of the $\eta'$ meson, and
study their behavior in the presence of mass splitting in
the quark sector. We find that even for realistic values of the quark masses
interesting
effects persist as $\theta$ is varied from $0$ to $2 \pi$.
Finally, we discuss the implications of these phase transitions
on axion models, in which $\theta$ is a dynamical variable.
In particular, we investigate the internal hadronic structure of axion domain
walls (previously studied by Huang and Sikivie \cite{HS}),
and their coupling to thermal degrees of freedom in the early universe.
We conclude that the domain walls can efficiently dissipate their kinetic
energy and therefore contract away very rapidly
after the QCD phase transition.

\vspace{-0.5cm}

\section{$N$ Degenerate Quark Flavors}

As a first example we consider QCD with $N$ quark flavors and a non-zero
$\theta$ angle
\begin{equation}
\label{QCD}
{\cal L} = - {1 \over 4 g^2} G^{\mu \nu} G_{\mu \nu} + \bar{\Psi_i} ( i
\gamma^\mu
D_{\mu}  - M)_{ij} \Psi_j
+ {\theta \over 32 \pi^2} \epsilon_{\mu \nu \alpha \beta} G^{\mu \nu} G^{\alpha
\beta}
\end{equation}
where $i$ runs over the $N$ flavors, and we shall take $M = m {\cal I}$.
The bare Lagrangian has an $SU(N)_L \otimes
SU(N)_R \otimes U(1)_V$ global flavor symmetry weakly broken by the
small current quark masses. The axial U(1) symmetry of the
classical theory is anomalous in the quantum theory. Using the chiral Ward
Identity
associated with the $U(1)_A$ symmetry we may rotate the $\theta$ term on to
the quark mass matrix which becomes $M_\theta = M e^{i \theta/N}$.

We shall assume that for all $\theta$
the theory confines  and breaks the chiral symmetry of the quarks to the vector
subgroup. At low energies the theory may then be described by the standard
$SU(N)_L \otimes SU(N)_R$ chiral Lagrangian
\beq
\label{L}
{\cal L} ~=~ \frac{F^2}{4} tr( \partial_{\mu} U^{\dagger}
\partial^{\mu} U ) + \Sigma Re tr(M_{\theta} U^{\dagger})~,
\eeq
where $U(x) = exp( 2 i \pi^a T^a / F)$ is an
$SU(N)$ matrix.
Taking a derivative with respect
to $M_{ij}$ in (\ref{QCD}) and (\ref{L}) we note that
\beq
\langle \bar{\Psi_i} \Psi_j \rangle =
\Sigma \langle U^{\dagger}_{ji} \rangle  ~,
\eeq
so $U$ can be directly related to condensates of quark bilinears.

The behavior of the minimum of the potential with changing $\theta$
is easily determined when $N=2$ since
the trace of a general $SU(2)$ matrix is real and
the potential reduces to
\beq
\label{pot2}
\epsilon_0 ~=~ -  m \Sigma~ cos ( \theta / 2) ~Re tr U
{}~\equiv~ - m \Sigma~ cos ( \theta / 2) ~u~,
\eeq
where we have defined $u \equiv Re tr U $. The variable
$u$ varies between $-1$ and $1$, and for
$cos( \theta / 2) \neq 0$ the potential energy is minimized
at one of those values. These vacua correspond to
$U$ matrices
\beq
\label{vac2}
U ~=~ \pm \pmatrix { 1 & 0 \cr 0 & 1}~.
\eeq
When $\theta = ({\rm odd ~integer}) \cdot \pi$ the
potential is exactly zero (at lowest order in the potential)
and all $U$'s are exactly
degenerate. It is easy to see that at these values of
$\theta$ there is a first order phase transition with
the vacuum jumping discontinuously between the two
configurations (\ref{vac2}).
We show the form of the potential  V graphically in Figure 1 for
varying $\theta$ and for the set of configurations
$U = {\rm diag} (e^{i\alpha}, e^{-i \alpha})$
(non-zero $\alpha$ corresponds to a $\pi^0$ condensate).
The vertical
axis in the figure is measured in units of $\Sigma m$.
The first
order phase transition at $\theta = \pi$ is clear.

Creutz \cite{Creutz} has investigated the global
behavior for general $N$.
In this case there are $N$ distinct minima  of the form
\beq
\label{Nvac}
U = e^{- 2 n  \pi i /N} {\cal I} \hspace{1cm}  n = 1,2,...N~~~,
\eeq
occurring for the
special values of $\theta$
\beq \theta_n = 2 n \pi \hspace{1cm} n=1,2,...N~~~~.  \eeq
These minima are connected by N first order phase transitions
at $\theta =$ (odd integer) $\pi$. We note that the minima correspond to the
vector flavor symmetry preserving diagonal SU(N) matrices.
The $U(1)_A$ symmetry provides
a simple understanding of the origin of the $N$ vacua
in (\ref{Nvac}), which
are characterized by bilinear quark condensates.
The $U(1)_A$ symmetry of (\ref{QCD}) is broken by the anomaly to
a $Z_{2N}$ discrete symmetry acting on the quark fields.
Therefore, discrete $Z_{N}$ transformations
on  the quark condensates
relate equivalent vacua of the theory.

$\left. \right.$  \hspace{-0.6in}\ifig\prtbdiag{}
{\epsfxsize9.5truecm\epsfbox{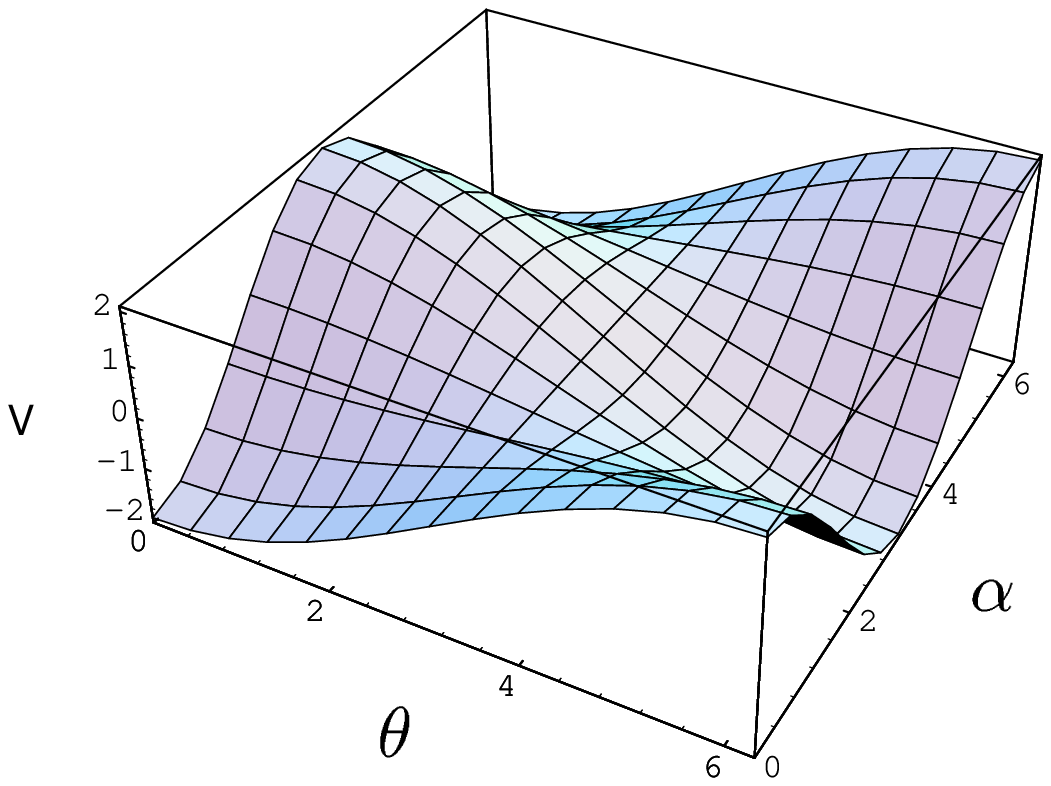}}  \vspace{-5.6cm} \\

\begin{center} Figure1: Potential, scaled by $\Sigma m$, for $U =
{\rm diag} (e^{i\alpha}, e^{-i \alpha})$
 and varying $\theta$ in  the degenerate
two flavor case. \end{center}

There is an important subtlety to discuss at this point.
To include the effects of
the $\theta$ angle in the chiral Lagrangian
we have made a $U(1)_A$ rotation on the
quark fields and moved the $\theta$ dependence into the quark mass
matrix. To put the low energy Lagrangians of bare theories with
different $\theta$ angles into the form of (\ref{L}),  {\it different}
axial rotations have been made in each case. Before comparing
the quark condensates in different models we must undo (or compensate for)
this  axial rotation.
We denote the matrix of condensates in the original basis of (\ref{QCD})
as $U_{0} = e^{ - i \theta /N} U $.
In the original basis the overall
phase of the quark condensate
changes smoothly  with $\theta$ and a discontinuous jump in the
$\pi^0$ condensate occurs at $\theta = \pi$.
This is shown in Figure 2 for the case of $N=2$,  where the
$U_{0}$  is parametrized as
\beq
\label{Uparam}
U_0 = e^{ - i \phi/2}  \left( \begin{array}{cc}  e^{i \alpha}
& 0 \\ 0 & e^{-i \alpha} \end{array} \right)~.
\eeq
After including the compensatory axial rotation, the
vacua at
$\theta = 0, 2 \pi$ are now identical, and given by
\beq
U_{0} ~=~ \pmatrix { 1 & 0 \cr 0 & 1}~.
\eeq
Similarly, in the $N$ flavor case the compensatory rotations also
cancel the phase factors in (\ref{Nvac}). Thus in the
original quark basis the $N$ vacua found by Creutz
are actually the same vacuum and correspond to $U_0 = {\cal I}$.

\vspace{1cm}

$\left. \right.$  \hspace{-0.3in}\ifig\prtbdiag{}
{\epsfxsize9.5truecm\epsfbox{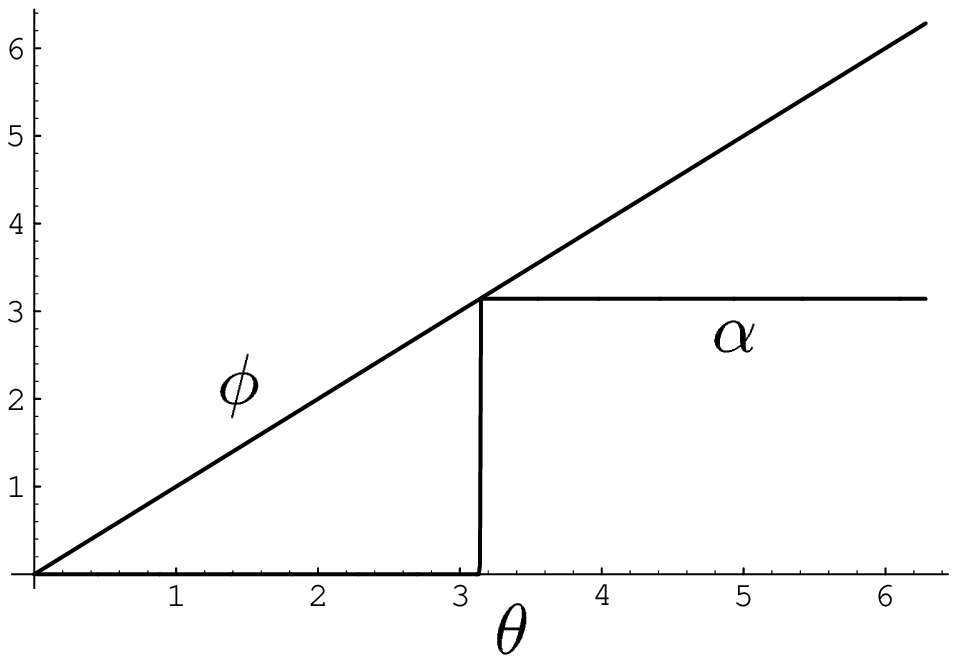}} \vspace{-1cm}
\begin{center} Figure 2: Phases of the vacuum $U_0$ matrix
as a function of $\theta$ in the degenerate two flavor case. \end{center}
\vspace{1cm}

In the two flavor case,
for $\theta = \pi \pm \epsilon$, the (compensated) minima are respectively
\beq
U_0 (\pi + \epsilon) = \left( \begin{array}{cc}  i & 0 \\ 0 & i \end{array}
\right), \hspace{1cm}
U_0 ( \pi - \epsilon) = \left( \begin{array}{cc}  - i & 0 \\ 0 & - i
\end{array} \right)
\eeq
Since they have an associated phase they violate CP. This model realizes the
possibility discussed by Witten that at $\theta = \pi$
there may be spontaneous CP violation by the occurence of
two CP violating vacua.
(CP is not explicitly violated when $\theta = \pi$
since physics is invariant to $2 \pi$ shifts in $\theta$, and CP acts by
changing the sign of any phase in the theory, $\theta = \pm \pi$
are equivalent.)

\section{Large $N_c$}

To better understand the physical implications of the vacuum dependence
on $\theta$ it is helpful to consider the problem in an $SU(N_c)$ gauge theory
at
large $N_c$. In this context the phase of $U$ can be identified with a
condensate in the $\eta'$ field.
It is well known that the anomaly in the $U(1)_A$ symmetry is the
result of the quark triangle graph which is suppressed at leading order in
$1/N_c$.
Thus at large $N_c$ the full $U(N)_L \otimes U(N)_R$ global symmetry on the
fermions is restored. The fermion condensates break the symmetry to $U(N)_V$
and there are $N^2$ goldstone bosons with the new state being identified with
the $\eta'$.  At ${\cal O}(1/N_c)$ in the $N_c$ expansion the
$\eta'$ acquires a small
mass. If for a given large value of $N_c$ we take the quark masses to zero
then this subleading term can dominate the $\eta'$ mass.  In \cite{eta}
it was argued that this is the limit most appropriate to realistic QCD.
We can hope therefore to
obtain some insight into the transition described above by looking to the large
$N_c$
limit.

The effective theory at large $N_c$ is \cite{Witten}
\beq
\label{LN}
{\cal L} ~=~ \frac{F^2}{4} tr( \partial_{\mu} U^{\dagger}
\partial^{\mu} U) + \Sigma Re tr(M U^{\dagger})~ - {
\tau \over 2} (\phi-\theta)^2.
\eeq
Here $U$ contains the $N^2$ goldstone degrees of freedom, and
is no longer restricted to unit determinant. The additional phase degree of
freedom correponds to the $\eta'$ meson, with

$$det U(x) = e^{- i \phi (x)}$$
and
\beq
\phi(x) = - {\sqrt{2 N} \over F} \eta'~.
\eeq
To obtain (\ref{LN}) we have performed an axial rotation on the bare Lagrangian
to
move the $\theta$ dependence into the mass matrix, then
undone the rotation by absorbing $\theta$ with a shift in the $\eta'$ field.
This
redefinition of the phase of $U$ corresponds exactly to the previously
discussed subtlety of
ensuring that we compare theories with different $\theta$ in the same quark
basis.
It is important to note that the final term in (\ref{LN}) is only the first
term
in an expansion in $\phi-\theta$  of a function that is $2\pi$ periodic.

Restricting our attention again to the case $N=2$ the potential
of the large $N_c$ theory is
\beq
\label{Npot}
V = -\Sigma Re tr(M U^{\dagger})~ + {
\tau \over 2} (\phi-\theta)^2~.
\eeq
We are interested in  the case $\tau / \Sigma m \rightarrow \infty$, where the
$\eta'$ is
very heavy, and we expect to recover the analysis of the
$SU(2)$ chiral Lagrangian above. The second term in the potential dominates
and is minimized by $\phi = \theta$. As $\theta$ changes between 0 and $2 \pi$
the phase of the mass matrix, which
is the $\eta'$ condensate, behaves as in Figure 2.
An ambiguity in the description arises here because a $\pi^0$ condensate
corresponding to $\alpha = \pi$ has the same effect as an $\eta'$
condensate of $\phi = 2 \pi$ on the matrix $U$ in (\ref{Uparam}).
This ambiguity is resolved in the presence of
mass splittings, as the discontinuities are smoothed,
and one can resolve the different behaviors of the $\pi^0$
and $\eta'$ condensates. We discuss the effect of mass splittings
below.

\section{Mass Splitting}

Returning to the $SU(N)$ chiral Lagrangian we may now consider the effects of
splitting the quark masses.  The potential is then
\beq
\label{split}
V = -\Sigma   \sum_i^{N-1} m_i \cos(\alpha_i - \theta/N) -
\Sigma m_N \cos(\theta / N + \sum_i^{N-1} \alpha_i)
\eeq
As before we have made
an equal axial rotation on each quark flavor to shift the $\theta$ angle
onto the quark mass matrix. The minima are of the form
\beq
\langle U \rangle = diag \left( e^{i\alpha_1}, e^{i\alpha_2},~...~,
e^{i\alpha_{N-1}}~,~
e^{-i\sum_i^{N-1} \alpha_i} \right)
\eeq
If we take  $m_N/m_i \rightarrow \infty$ the last term in (\ref{split})
dominates and the heavy quark
condensate acquires a phase $\sum_i^{N-1} \alpha_i = \theta/ N$.
In other words the phase of
the $N$th diagonal element of $UM_\theta$ goes to zero corresponding
to decoupling of the heavy quark and the problem reduces
to that  in which the $\theta$ angle has been rotated onto just the $N-1$ light
quarks. This is why the strong CP problem cannot be solved by simply
rotating the $\theta$ angle solely onto a heavy
quark such as the top. In nature
the charm, bottom and top quarks are sufficiently heavy that they essentially
decouple from the $\theta$ dependence of the vacuum.

For intermediate mass splittings we must minimize (\ref{split}).
For $N=2$ this is most easily done numerically.
Minimizing  (\ref{split}) and rotating the resulting $U$ back to
the quark basis of (\ref{QCD}) we have
\beq
U_0 = e^{-i \theta/2}  \left( \begin{array}{cc}  e^{i \alpha}
& 0 \\ 0 & e^{-i \alpha} \end{array} \right)~.
\eeq
We show the phase $\alpha$  that minimizes the potential at varying
$\theta$ for increasing mass splitting in Figure 3.
The first order phase transition is smoothed out by
the mass splitting and  as $m_d/m_u \rightarrow \infty$
we recover the one flavor case for which there is a single vacuum.

\vspace{-2cm}
$\left. \right.$  \hspace{-0.3in}\ifig\prtbdiag{}
{\epsfxsize9.5truecm\epsfbox{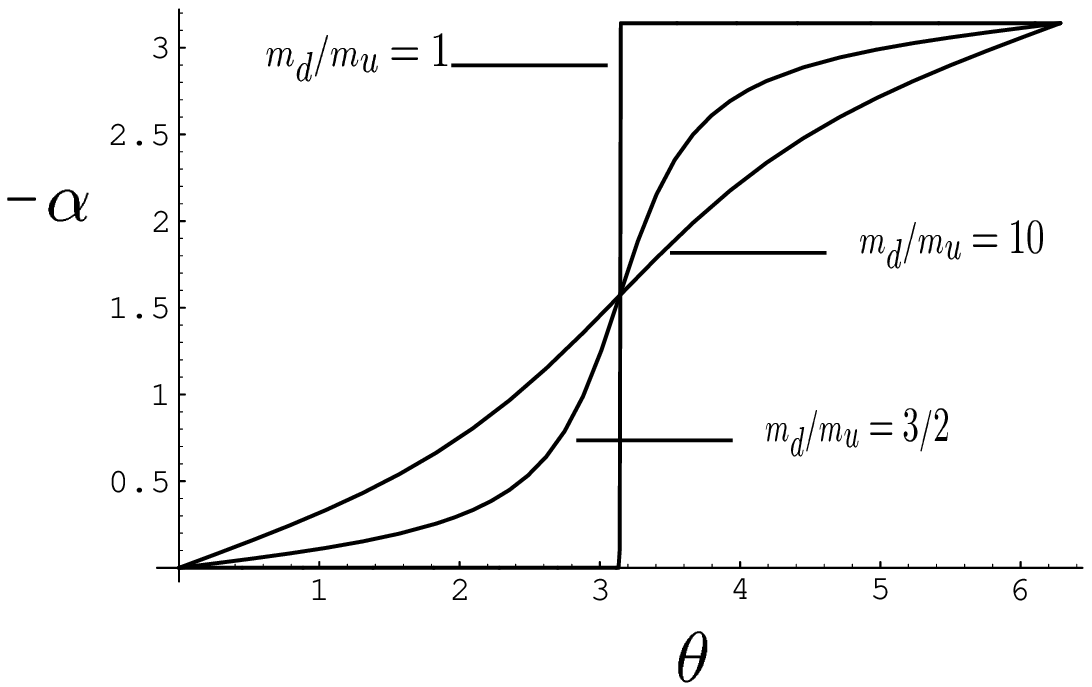}}  \vspace{-2cm} \\
\begin{center} Figure 3: The phase $\alpha$ vs $\theta$ in the presence of mass
splitting. \end{center}

With mass splittings there is only a single CP conserving minimum at
$\theta = \pi$:
\beq
U_0 =  \left( \begin{array}{cc}  -1
& 0 \\ 0 & 1 \end{array} \right)~~.
\eeq
This minimum violates the vector symmetry of the model, however,
this is just a reflection of the explicit breaking of that symmetry in
the non-degeneracy of the quark masses.
Furthermore, in this case one can easily see by examining the
form of $U$ that both the
$\eta'$ and the $\pi^0$  condensates are changing with $\theta$.

The three flavor case is more complicated. The minimization equations
reduce to (with $U =diag(e^{i \alpha}, e^{i \beta}, e^{-i (\alpha + \beta)}$))
\beq \label{mins}
m_u \sin(\alpha - \theta/3) = m_d \sin (\beta-\theta/3) = - ~
m_s \sin(\alpha + \beta + \theta/3)
\eeq
The behavior is most easily studied by displaying the solutions
at $\theta = \pi$, where phase transitions are likely to occur.
There are several types of solutions to (\ref{mins}).
Trivial solutions occur when the sine functions
are all simultaneously zero. For realistic quark masses
($m_u \sim 5 \mev$, $m_d \sim 10 \mev$, $m_s \sim 150 \mev$) one of
these solutions is the global  (and only)
minimum corresponding to the heavy strange
quark having decoupled and the minimum
closely resembles the $N=2$ case with $m_u \neq m_d$
(See Figure 3).
Non-trivial solutions were studied previously by
Witten \cite{Witten}, and occur when the quark masses
satisfy
\beq
\label{region}
m_u m_d \geq m_s |  m_d  - m_u | ~,
\eeq
which includes the degenerate mass limit discussed above.
They satisfy
\beq
1 - \cos(\alpha + \beta + \pi/3) = { m_u^2 m_d^2 - m_s^2(m_d-m_u)^2 \over
2 m_s^2 m_u m_d} ~.
\eeq
These latter solutions exhibit spontaneous
CP violation. In this case the trivial solutions discussed above
become points of inflection.
Witten's solutions are probably excluded for realistic values
of the light quark masses.

As an explicit example we have examined the behavior with varying
$\theta$ of the Large-$N_c$ effective Lagrangian.
We used the physical $\eta, \eta'$ masses to determine
$\tau / \Sigma = 200 \mev$ and $m_s = 150 \mev$.
With such large values of $\tau$ and $m_s$ the $\tau$ and
$m_s$ dependence of the solutions effectively decouples,
and we are left with the behavior of the $SU(2)$ chiral Lagrangian.
For example, taking $m_u = m_d = 7 \mev$, we recover the
first order transition of the degenerate two flavor case.
Minimization in the $SU(3)$ degrees of freedom in
$U=e^{-i \phi/3} diag(e^{i (\alpha +\beta)}, e^{i (-\alpha +\beta)}, e^{-i (2
\beta)})$
leads to  discontinuities in the $\pi^0$, $\eta$ and $\eta'$
condensates at $\theta = \pi$ (Figures 4,5).

\vspace{1cm}

$\left. \right.$  \hspace{-0.3in}\ifig\prtbdiag{}
{\epsfxsize8truecm\epsfbox{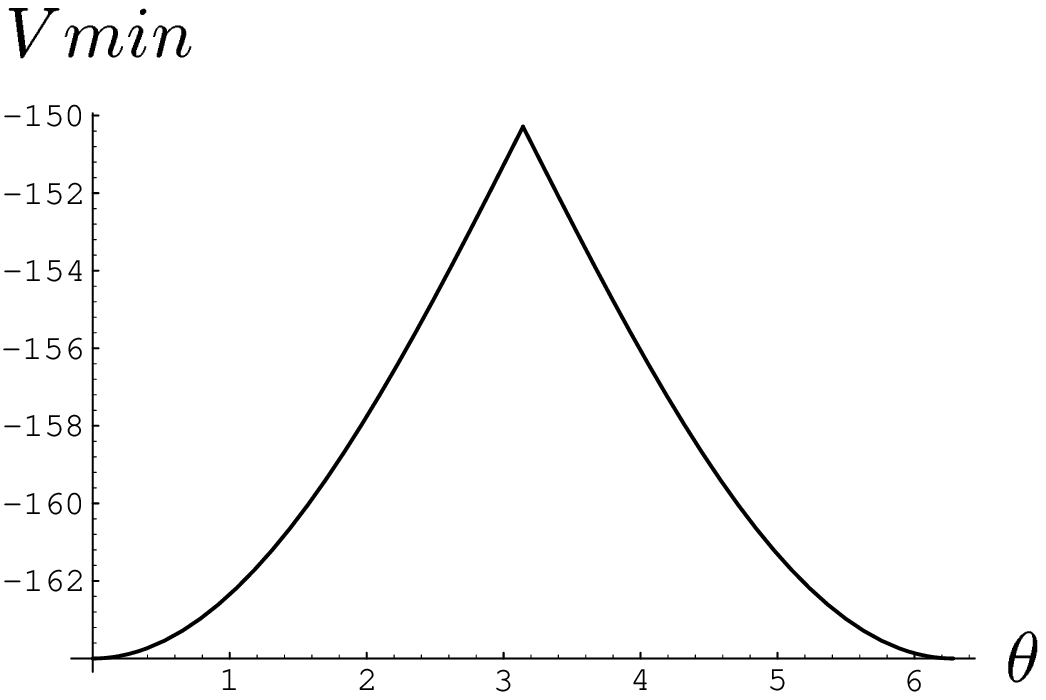}} \vspace{-1cm}
\begin{center} Figure 4:
The value of the potential at the minimum as a function of $\theta$  for the
case
of degenerate $u$ and $d$ quarks, $m = 7 \mev$,  $m_s = 150 \mev$,
$\tau / \Sigma = 200 \mev$.
 \end{center}
\vspace{1cm}

\vspace{1cm}

$\left. \right.$  \hspace{-0.3in}\ifig\prtbdiag{}
{\epsfxsize17truecm\epsfbox{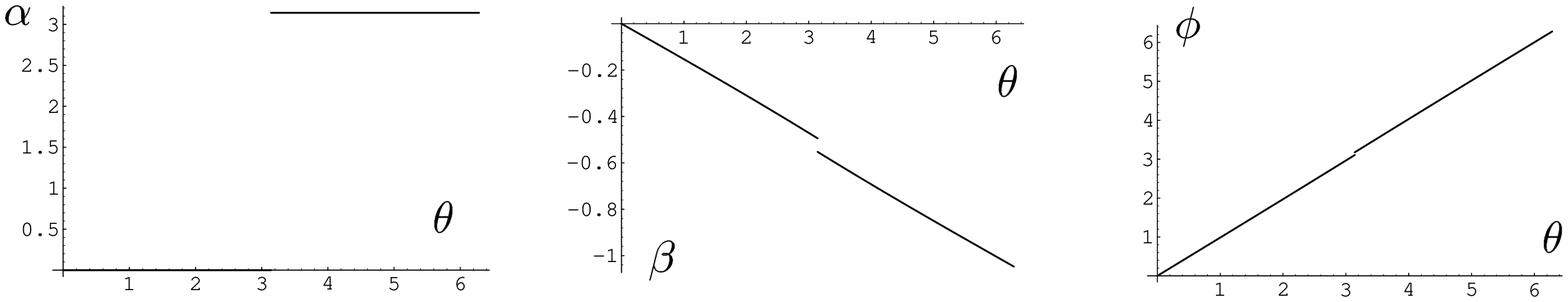}} \vspace{-1cm}
\begin{center} Figure 5:
The condensates $\alpha \propto \pi^0$, $\beta \propto \eta$,
 $\phi \propto \eta'$  as functions of $\theta$.
 \end{center}
\vspace{1cm}

If one moves away from $\theta = \pi$ one of the two solutions of the
minimization equation~(\ref{mins}) becomes the global minimum, whereas
the other solution is a local minimum. By crossing $\theta = \pi$ the
global minimum switches between the two branches. This description is
valid in the vicinity of $\theta = \pi$. For $\mid \theta - \pi \mid >
m_u/m_s$ the second (local) minimum disappears and there is only one
(global) minimum. The situation is similar for a small mass splitting
between $m_u$ and $m_d$, although Witten's condition~(\ref{region}) is
modified for $\theta \neq \pi$. Finally, for large mass splittings
there is only one minimum of the potential for all values of $\theta$
(see also \cite{Creutz}).

\newpage

\section{Axion Cosmology}

The results described above have some
interesting implications for axion cosmology,
particularly for non-trivial configurations like
axion domain walls (ADWs). The internal structure of the
ADW was investigated previously
by Huang and Sikivie \cite{HS}, who were the
first to notice its internal $\pi^0$ structure.
In this section we emphasize the additional
$\eta, \eta'$ structure of the wall and the
nature of the coupling of an ADW to thermal
degrees of freedom below the QCD phase
transition.

In axion models \cite{CP}
the role of the $\theta$ parameter is
played by a dynamical axion field $a$,
whose potential energy is minimized at
$\theta$ very close to zero, thus providing
a solution to the strong CP problem.
The models must exhibit an anomalous
Peccei-Quinn symmetry, $U(1)_{PQ}$,
at high energies, which is spontaneously
broken at some scale $f_{PQ}$.

Because of the spontaneous breaking of
a $U(1)$ symmetry, the cosmology of these
models is somewhat involved. At temperatures below
$f_{PQ}$, global cosmological strings are formed,
which persist until the QCD chiral phase transition
(unless inflation occurs with reheat temperature
below $f_{PQ}$).
At the QCD phase transition the degeneracy corresponding to
different values of $\langle a \rangle / f_{PQ} \sim  \theta$
is lifted, with the energy minimized at
$\theta = 2 \pi n ~,~~ n \in {\cal Z}$ in the simplest
(N=1) axion models.  Since the axion field undergoes
a `winding' about the axis of the string, after the chiral phase
transition  the axion strings become connected
by domain walls in which the axion field interpolates
from $\theta = 0$ to $\theta = 2 \pi$ over a
lengthscale $m_a^{-1}$. Our earlier results imply that
there are also domain walls in the
$\pi^0$, $\eta$ and $\eta'$ condensates
across the ADW. We note that
the physical axion and
Goldstone fields are obtained by diagonalizing the
low energy mass matrix, see for example \cite{mass}.

The subsequent evolution of a network
of axion strings connected by domain walls
was studied in \cite{SS}, with the conclusion that
the ADWs could persist for many horizon times
and eventually even affect structure formation.
This conclusion relies on the assumption that the
ADW is extremely weakly coupled to the thermal
background particles\footnote{It also assumes that
an ADW does not rapidly destroy itself by self-intersection.}.
In \cite{HS} this assumption was justified by detailed
calculations. However, only the $\pi^0$ component
of the ADW was considered in \cite{HS}, and in particular
only the derivative couplings of the $\pi^0$ to nucleons.
The derivative couplings lead to effects which are
suppressed by the spatial derivative of the wall, which
is $\sim m_a$ and hence very small.

There are additional interactions of the ADW with the
thermal background which lead us to believe that, in fact,
the ADW is strongly coupled and can rapidly dissipate
excess energy.  The relevant
couplings are due to

\begin{itemize}

{\item The internal $\eta, \eta'$ structure of the wall, which
couples strongly to all hadronic thermal modes. In particular,
the $\eta'$ interactions can be estimated from the large-$N_c$
effective Lagrangian, yielding interactions such as
$n$-point $\eta'$ vertices which are suppressed at most
by powers of $1/N_c$, but are otherwise of typical
strong interaction size. }

{\item The non-derivative couplings of the $\pi^0$ domain
wall, which are suppressed by powers of the light
quark masses (explicit chiral symmetry breaking) but which are
still non-negligible. For example there are four $\pi$ interactions
with coefficient $\sim \Sigma m / f_{\pi}^4$. }

\end{itemize}

Due to the interactions mentioned above, the
ADW will appear as a potential well of depth
$\Lambda_{QCD}$
to the thermal hadrons at the QCD phase transition,
which we expect to scatter with probability of order one.
The ADW will contract
at a rate \cite{HS}
\begin{equation}
\label{dis}
{ d E \over d t } \sim  \sigma  {d l^2 \over d t } \sim - \rho l^2
\end{equation}
where $l$ is the size of the ADW,
$\sigma \sim f_{PQ} f_{\pi} m_{\pi}$  is
the surface tension of the domain wall
and $\rho$ the hadron
energy density. We will focus on the contribution to (\ref{dis})
from pions, because they are likely to have the highest energy
density after the QCD phase transition.
Since the quark condensate approaches zero at the chiral
phase transition, we expect that
the effective masses of the pions should be reduced.
We therefore estimate the
pion energy density to be given roughly by the standard relativistic
formula for massless ($m << T$) particles
\begin{equation}
\rho_{\pi} = {\pi^2 \over 30} T^4 \sim f_{\pi}^4~.
\end{equation}
This should be compared with the energy density of
axions \cite{HS} which is much smaller: %
\begin{equation}
\rho_a \sim 10^{-9} f_{\pi}^2 m_{\pi}^2~.
\end{equation}
The  contraction rate of the ADW from interactions with the $\pi$
and $\eta'$
thermal background is therefore very large
\beq
\label{dldt}
{dl \over dt} \sim {f_{\pi}^2 \over  f_{PQ} } l
{}~.
\eeq
The rate of shrinkage in (\ref{dldt}) is to be compared
with the horizon doubling time $\sim M_{P} / f_{\pi}^2$.
Since the Planck scale $M_P >> f_{PQ}$,
we expect the ADW to contract away
in a relatively short cosmological timescale, even if
the effects of self-intersections are ignored.

\vskip 1in
\newpage
\centerline{\bf Acknowledgements}
\vskip 0.1in
The authors would like to thank Pierre Sikivie for
useful discussions.
This work was supported under
DOE contract DE-AC02-ERU3075. A.N. also acknowledges
the support of Schweizerischer Nationalfonds.


\end{document}